\documentclass[12pt,a4wide]{article} 
\usepackage{epsfig}
\begin{document} 

\title{\bf Ground-state energy of a high-density electron gas in a strong magnetic field}
\author{M.Steinberg, J.Ortner\\ \\Humboldt-Universit\"at zu Berlin \\Invalidenstr.110, 10115 Berlin}  
\date{}

\maketitle

\begin{abstract}
The high-density electron gas in a strong magnetic field $B$ and at zero temperature is investigated. The quantum strong-field limit is considered in which only the lowest Landau level is occupied. It is shown that the perturbation series of the ground-state energy can be represented in analogy to the Gell-Mann Brueckner expression of the ground-state energy of the field-free electron gas. The role of the expansion parameter is taken by $r_B= (2/3 \pi^2)\,(B/m^2)\,(\hbar r_S /e)^3$ instead of the field-free Gell-Mann Brueckner parameter $r_s$. 
\end{abstract} 

\thispagestyle{empty}

\section{INTRODUCTION}

The interacting electron gas is a widely used model in condensed matter, chemistry, and astrophysics and has been studied in great detail. Many investigations have focussed on the calculation of the ground-state energy. Gell-Mann and Brueckner \cite{Gell-Mann&Brueckner} derived an exact formula for the ground-state energy of a high-density electron gas. Their method will also be employed in this work. 

In contrast to the field-free case, the electron gas in an externally applied magnetic field has received much less attention. So far most of the works have concentrated on studying the ground-state properties in random-phase approximation \cite{Gl76}.   

Clearly, it is of general interest to find analytical expressions for at least some limiting cases. In doing so, we concentrate in our calculation on the strong-field limit in which only the lowest Landau level is assumed to be occupied and all electrons are aligned antiparallel to the magnetic field. In this case the energy associated with a Landau level exceeds the Fermi energy. This regime has been investigated by Horing {\sl et al.} \cite{Horing&Danz&Glasser}, Isihara and Tsai \cite{Isihara}, Keldysh and Onishchenko \cite{KeOn76}, and recently by Skudlarski and Vignale \cite{SkVi93}, and by Steinberg and Ortner \cite{StOr98}. The latter have obtained an asymptotically exact result by deriving an analogous expansion of the ground state energy as it was established by Gell-Mann Brueckner. Within this calculation the result derived by Horing {\sl et al.} \cite{Horing&Danz&Glasser} was found to be the leading term in the expansion.  
Keldysh {\sl et al.} \cite{KeOn76} considered the limiting case of a small filling factor of the lowest Landau level. Additionally, they performed an expansion of the RPA correlation energy in inverse powers of the Wigner-Seitz radius $r_s$ and derived an result that becomes exact in the limit of small densities, i.e., $r_s>1$. Strictly speaking, the RPA is not valid in this regime. However, at metallic densities the RPA is considered to be a reasonable approximation of the correlation energy.\par
Physically, the strong-field limit may be achieved in laboratory plasmas, such as semiconductors, but one also expects these conditions on the surface of neutron stars. Strong magnetic fields are predicted to be observable in laser induced plasmas \cite{PuMtV96}. The results for the ground-state energy in the high density approximation may also be used as a starting point for the construction of the local-density approximation for the density-functional theory. \par 

Unlike in the field-free electron gas, the Fermi energy is no longer a function of the Wigner-Seitz radius $r_s$ alone, but also of the magnetic field. Therefore, it is convenient to introduce the new parameters \cite{StOr98}
\begin{equation}
\label{1} r_B = \frac{1}{\pi a_B k_F} =\frac{2}{3\pi^2} \alpha^2 r_s^3 \, \,
, \, \hspace{0.5cm}\, \hspace{0.5cm} t=\frac{\epsilon_F}{\hbar \omega_c}=\frac{9\pi^2}{8} \frac{1}{\alpha^6 r_s^6} \, ,
\end{equation} 
where $\omega_c=eB/m$ is the cyclotron frequency, $\alpha=a_B/l_B$ is the ratio
of the Bohr radius and the magnetic length $l_B=\sqrt{\hbar/(eB)}$. At strong magnetic fields $r_B$ takes the role of the expansion parameter. The second parameter $t$ may be regarded as a filling parameter. In the quantum strong-field limit it satisfies the condition $t\leq1$. Now the ground-state energy (per Rydberg and electron) of a high-density electron gas in the quantum strong-field limit is of the following form 

\begin{equation}
\label{2} \epsilon_{g} = \frac{1}{3\pi^2} \frac{1}{r_B^2}+\frac{A(t)}{r_B}+B(
t) \ln(r_B) + C(t)+ {\rm terms \, \, that \, \, vanish \, \,  as \, \, r_B \rightarrow 0} \, .
\end{equation}

This gives the asymptotically exact ground-state energy in the limit $r_B \rightarrow 0$. The coefficients $A(t)$, $B(t)$, and $C(t)$ in this expansion depend on the filling factor $t$. The leading term in (\ref{2}) is the kinetic energy, while the second term describes the first-order exchange effects. The sum of all remaining contributions is known as the correlation energy. Only the RPA correlation energy contributes to $B(t)$ in the high-density limit, i.e. $r_B \rightarrow 0$. To obtain $C(t)$, we must calculate the second-order exchange diagram in addition to the sum of all ring diagramms. 

In an earlier work \cite{StOr98} the authors have given analytic expressions for the constants $A(t)$, $B(t)$, and $C(t)$ in the limit of small filling factors $t \rightarrow 0$. In this regime all electrons occupy the lowest Landau level and transitions to higher levels are negelected. We will now generalize these results and consider the case of arbitrary filling factors by taking into account transition to higher levels. With that, one can improve the quality of the convergence of the series. The validity domain of this expansion in the $r_s$-$t$ plane is schematically shown in Figure 1.

\begin{figure}[h]
\centerline{\epsfig {figure=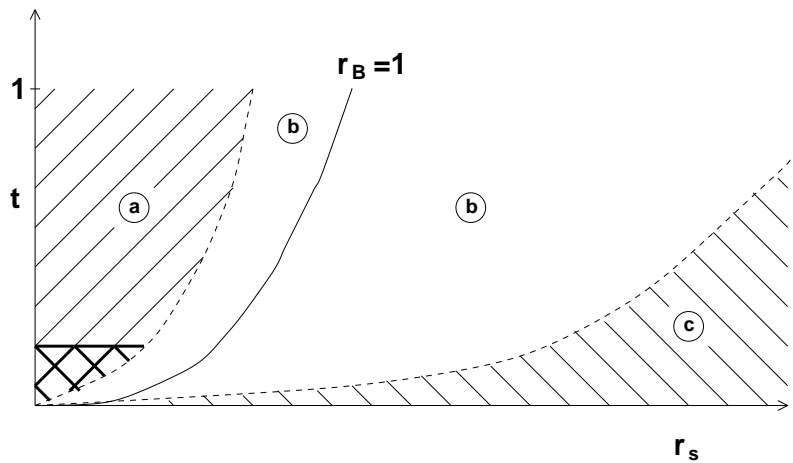,width=10.4cm,angle=0}}
\caption{\small \sl Sketch of the validity domain of the various expansions in the $r_s-t$ plane (taken from ref. \cite{StOr98}): (a) high-density region [the cross-hatched area at the bottom shows the validity domain of the analytic results for the asymptotic behavior $t \rightarrow 0$ of the coefficients  $A(t)$, $B(t)$ and $C(t)$ as given in \cite{StOr98}.], (b) intermediate-density region, and (c) low-density region. }
\end{figure}

\section{HARTREE-FOCK AND CORRELATION ENERGY}

The dominant interaction contribution to the internal energy of a high-density electron gas comes from the first-order exchange term. It was first calculated by Danz and Glasser \cite{Danz&Glasser} in the quantum strong-field limit. They derived an integral representation for $A(t)$, which may be expressed in terms of Meijer's $G$-function. By extending their calculation we have found the following series 

\begin{equation}
\label{3} A(t) =  - \frac{1}{\pi^2} \left(3-{\bf C} -\ln(4t) \right) -  \frac{1}{\pi^2} \sum_{s=1}^\infty \frac{(4t)^s}{s!(2s+1)(s+1)} \left( \psi(s+1)-\ln(4t)+\frac{4s+3}{(2s+1)(s+1)}\right)\, ,
\end{equation}
where ${\bf C}$ is Euler's constant ${\bf C} \approx 0.5772$. The first term determines the asymptotic behavior for small filling factors.   \par

\thispagestyle{empty}     

The starting point for the calculation of the correlation energy of a high-density electron gas is the determination of the polarization function in the random-phase approximation. 
An explicit expression for the polarisation function in the quantum strong-field limit was derived by Horing et al.\cite{Horing&Danz&Glasser}. Using these results and following the original work of Gell-Mann and Brueckner, we expand the polarization function in powers of the momentum transfer. In order to avoid divergencies, we introduce a cut-off in the following momentum integration. However, the final result will be independent of this cut-off procedure. After performing all momentum integrations, we find the exact result for the logarithmic contribution \cite{Horing&Danz&Glasser,StOr98} 

\begin{equation}
\label{4} B(t)=\frac{1}{16\pi^2t} \, .
\end{equation}  
Furthermore, we can extend this procedure to obtain the next term in this series, which will be independent of $r_B$. The constant of this expansion may be split into contributions coming from the ring approximation and from the exchange term according to $C(t)=C^{RPA}(t)+C^{ex}(t)$. In the following, we only consider the RPA contribtuion to $C(t)$, since even the numerical evaluation of $C^{ex}(t)$ is very complicated. In Ref.\cite{StOr98}, we have analytically found an expression for $C^{RPA}(t)$ at small values of $t$. Additionally, we have derived an integral representation for $C^{RPA}(t)$ at any filling factor, which we have now numerically integrated. The final expression for $C^{RPA}(t)$ may be obtained from a numerical fit to the results and is given by

\begin{equation}
\label{5} C^{RPA}(t)=-0.00633 \frac{\ln(t)}{t}-\frac{0.02706}{t}-0.16666 \ln (t)-0.59661+ at^b+ct^{1/2} \, ,
\end{equation}
with a=-2.23106, b=0.5677 and c=2.77534. The first three terms were analytically established in \cite{StOr98} and give the exact asymptotics at $t \rightarrow 0$. The remaining contributions are fitted in such a way that the total correlation energy, including the logarithmic term, is accurate within $1\%$.

\thispagestyle{empty}
\begin{figure}[h]
\centerline{\epsfig {figure=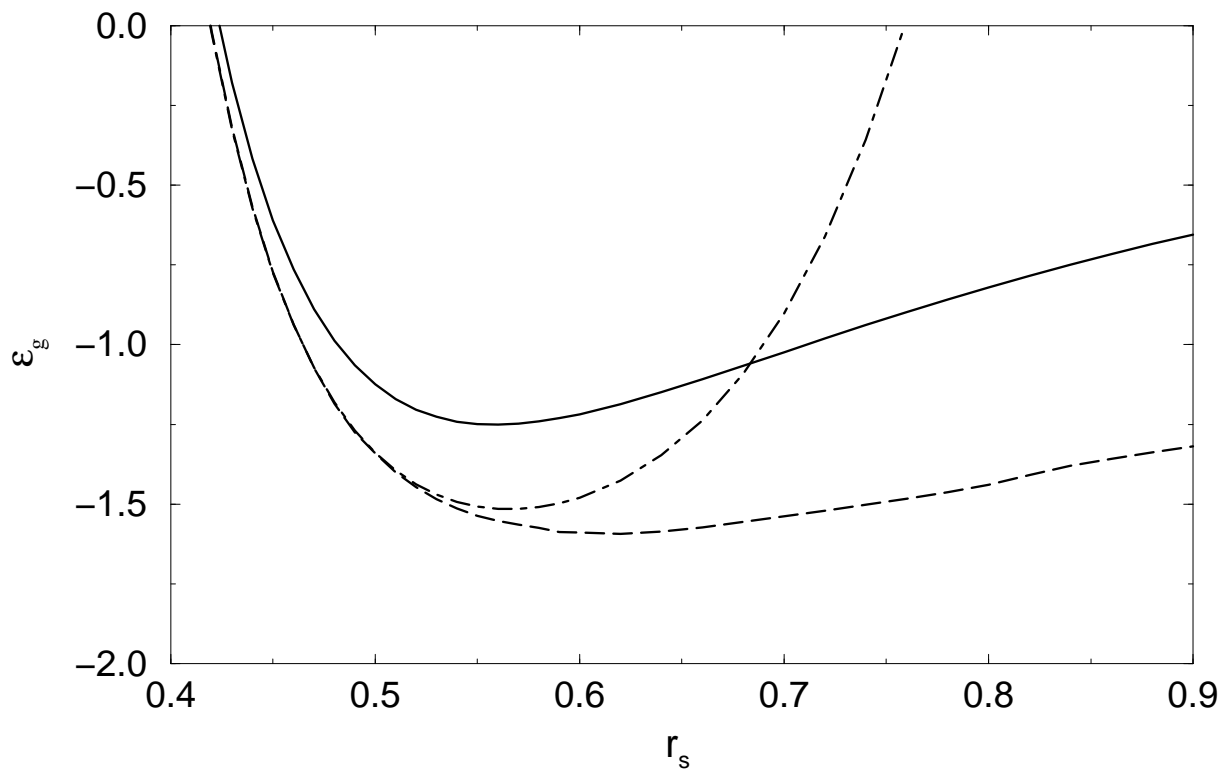,width=10.4cm,angle=-90}}
\caption{\small \sl Ground-state energy in Hartree-Fock approximation (solid line), in high-density approx. (dot-dashed line), and in the full RPA approx. (dashed line) for $\alpha=4.673$.}
\end{figure}


In Figure 2 we have plotted the ground-state energy in the Hartree-Fock and the high-density approximation (\ref{2}). For comparison we have included the numerical results for the full RPA ground-state energy. At lower densities ($r_s \geq 0.52$ at $\alpha=4.673$) the high-density ground-state energy deviates from the full RPA ground-state energy indicating the breakdown of the high-density expansion. At these densities the correlation energy should be calculated from an improved polarisation function including local-field corrections \cite{Ich92}.

\section[Acknowledgments]{Acknowledgments}
This work was supported by the Deutsche Forschungsgemeinschaft under grant\#Eb 126/5-1. We thank Werner Ebeling for useful discussions.


\begin{thebibliography}{99}        

\thispagestyle{empty}

\bibitem{Gell-Mann&Brueckner}
M.Gell-Mann, K.A.Brueckner, Phys. Rev. {\bf 106}, 364 (1957)

\bibitem{Gl76}
For an early review, see M.L.Glasser, Th. Chem. Advances and Perspectives {\bf 2}, 67 (1976)

\bibitem{Horing&Danz&Glasser}
N.J.Horing, R.W.Danz, and M.L.Glasser, Phys. Rev. A {\bf 6}, 2391 (1972)

\bibitem{Isihara}
A.Isihara, J.T.Tsai, Phys. kondens. Mater. {\bf 15}, 214 (1972)

\bibitem{KeOn76}
L.V.Keldysh, T.A.Onishchenko, Pis'ma Zh.Eksp.teor.Fiz. {\bf 24}, 70 (1976)

\bibitem{SkVi93}
P.Skudlarski, G.Vignale, Phys. Rev. B {\bf 48}, 8547 (1993)     

\bibitem{StOr98}
M. Steinberg, J. Ortner, Phys. Rev. B 58, 15 460 (1998); Erratum: {\bf 59}, 12693 (1999)

\thispagestyle{empty}

  

\bibitem{PuMtV96}
A.Pukhov, J.Meyer-ter-Vehn, PRL {\bf 76}, 3975 (1996)

\bibitem{Danz&Glasser}
R.W.Danz, M.L.Glasser, Phys. Rev. B {\bf 4}, 94 (1971)

\bibitem{Ich92} 
A survey on various approximations for obtaining local field corrections in the field free case may be found in: S.Ichimaru, Rev. Mod. Phys. {\bf 54}, 1017 (1992) 
\end{thebibliography}
\end{document}